\documentclass[3p,onecolumn,10pt,times]{elsarticle}

\usepackage[utf8]{inputenc}
\usepackage{amsmath,amssymb}
\usepackage{graphicx}
\usepackage{xcolor}
\usepackage{physics}
\usepackage{ifthen}
\usepackage{booktabs} 
\usepackage{multirow}
\usepackage{siunitx}
\usepackage{tikz-feynman}
\usepackage[]{changes}

\usepackage{tikz}
\usetikzlibrary[plotmarks,shapes,decorations.pathmorphing]

\tikzset{snake it/.style={decorate, decoration=snake}}

\usepackage[colorlinks=true, citecolor=red, linkcolor=blue, urlcolor=magenta]{hyperref}

\newcommand{\nopi}{\ensuremath{{}^{\pi\!\!\!/}\mathrm{EFT}}}

\newcommand{\Hd}{\ensuremath{^2\mathrm{H}}}
\newcommand{\Ht}{\ensuremath{^3\mathrm{H}}}
\newcommand{\Het}{\ensuremath{^3\mathrm{He}}}
\newcommand{\Hef}{\ensuremath{^4\mathrm{He}}}

\newcommand{\rms}{\ensuremath{\sqrt{\left<{r}_p^2\right>}}}

\newcommand{\figref}[1]{Fig.~\ref{#1}}
\newcommand{\secref}[1]{Sec.~\ref{#1}}
\newcommand{\tabref}[1]{Table~\ref{#1}}
\begin{document}
\begin{frontmatter}
\title{A practical approach to perturbative corrections to few-body observables}
\author[iitg]{Sourav Mondal}
\ead{sm206121110@iitg.ac.in}
\author[npi]{Martin Sch\"afer}
\ead{m.schafer@ujf.cas.cz}
\author[heb]{Mirko Bagnarol}
\author[heb]{Nir Barnea}
\author[iitg]{Udit Raha}
\author[uzb]{Johannes Kirscher}

\affiliation[iitg]{organization={Department of Physics, Indian Institute of Technology Guwahati}, city={Guwahati}, postcode={781039}, country={India}}

\affiliation[npi]{organization={Nuclear Physics Institute of the Czech Academy of Sciences}, city={Rez}, postcode={25068}, country={Czech Republic}}

\affiliation[heb]{organization={The Racah Institute of Physics}, addressline={The Hebrew University}, city={Jerusalem}, postcode={9190401}, country={Israel}}

\affiliation[uzb]{organization={New Uzbekistan University}, addressline={Movarounnahr Street 1}, city={Tashkent}, postcode={100000}, country={Uzbekistan}}

\begin{abstract}
We formulate two methods to facilitate the calculation of perturbative corrections to quantum few-body observables. Both techniques are designed for a numerical realization in combination with any tool that obtains either the entire spectrum or solely the eigenvalues of an operator corresponding to the observable of interest. We exemplify these methods in the context of the nuclear contact theory without pions (\nopi) and benchmark them in the deuteron channel with available analytical, field-theoretical calculations, as well as in the triton and 3-helium channels through earlier extractions within the dibaryon formalism, where in all three systems the point-proton root-mean-square charge radius (rms) was the perturbed observable of choice. Beyond these $A\leq3$ consistency and accuracy checks, we employ the numerical methods to predict the rms of the 4-helium nuclear ground state to assess three different ways of integrating the Coulomb interaction into~\nopi. By comparing the respective results at leading and next-to-leading order for 3- and 4-helium, we find that the uncertainty due to the strong, short-range interaction is significantly larger compared with that due to the long-range Coulomb interaction for both bound states with their different binding momenta. Thereby, we provide strong support for simplifying extractions of bound-state observables by shutting off any Coulomb interaction if the strong part of the potential is considered only up to first order in the effective range expansion.
\end{abstract}

\begin{keyword}
  perturbation \sep few-body nuclei \sep effective field theory \sep Coulomb interaction  \sep root-mean-square radius 
\end{keyword}

\end{frontmatter}

\section{Introduction}
Self-consistency of a physical theory sometimes demands the perturbative treatment of some of its parts. For instance, causality imposes
constraints on the interaction range of a non-relativistic theory that is required to describe a system with correlations over a certain 
spatial extent, the so-called Wigner bound~\cite{Wigner:1955zz}. For effective field theories (EFTs) this often means that a resummation 
of range corrections leads to renormalization problems by violating the bound \cite{phillips1997,Hammer:2010fw}. To bypass this issue and
thus maintain a clear hierarchy of EFT orders, the sub-leading EFT terms, including the range, are often treated perturbatively.

Counterintuitively, the practical extraction of perturbative corrections of a given order to few- and many-nucleon observables is often 
less straightforward than obtaining predictions through an infinite iteration of the perturbation in the Schr\"{o}dinger equation. For our specific interest in $A$-body systems described by a Hamiltonian with non-perturbative kinetic 
(\(\hat{T}\)) and interaction terms (\(\hat{V}^{(0)}\)) plus a relatively weak part (\(\alpha \hat{V}^{(1)}\)) that acts as a 
perturbation,
\begin{equation}\label{eq.ham}
\hat{H}=\hat{T}+\hat{V}^{(0)}+\alpha\hat{V}^{(1)} = \hat{H}^{(0)} +\alpha \hat{V}^{(1)}~,
\end{equation}
the technical hurdle can be understood in terms of textbook quantum mechanics as follows: At $0^{\text{th}}$ order ($\alpha=0$), 
the expectation value of an observable $\hat{O}$ is given by
\begin{equation}
O^{(0)}_n  = \mel**{\psi^{(0)}_n}{\hat{O}}{\psi^{(0)}_n}\,,
\end{equation}
where $\ket{\psi^{(0)}_n}$ is the $n^{\text{th}}$ eigen solution corresponding to the zeroth order Hamiltonian $H^{(0)}$. The 
leading perturbative correction to the expectation value $O^{(0)}_n$ is linear in the small parameter $\alpha$ and is given by
\begin{equation}\label{eq.prms}
\delta O^{(1)}_n = \mel**{\psi^{(1)}_n}{\hat{O}}{\psi^{(0)}_n}+\text{h.c.}\,,
\end{equation}
where it is the first order correction to the $n^{\text{th}}$ eigenstate, expressed via the infinite sum
\begin{equation}\label{eq.psi1}
    \ket{\psi^{(1)}_n}=\sum_{m \ne n}
    \frac{\mel{\psi^{(0)}_m}{\hat{V}^{(1)}}{\psi^{(0)}_n}}{E_n - E_m}\ket{\psi^{(0)}_m}\,,
\end{equation}
which often precludes rigorous perturbative calculations, as it incorporates contributions from all possible
eigenstates $\ket{\psi^{(0)}_m}$ together with their corresponding eigenenergies $E_m$. 

For the specific case of the energy, however, the problem simplifies. While the first-order perturbative correction to the energy can be calculated
as an expectation value $\mel**{\psi^{(0)}_n}{\hat{V}^{(1)}}{\psi^{(0)}_n}$, the infinite sum, \eqref{eq.psi1}, appears at second order.
Alternatively, these corrections can be extracted from a solution of the Schr\"{o}dinger equation with Hamiltonian
$\hat{H}$, \eqref{eq.ham}, and a polynomial expansion of $E_n(\alpha)$
in the small parameter $\alpha$. Such a procedure
was successfully applied in Refs.~~\cite{yang2016,Yang:2020pgi,Yang:2021vxa}, where perturbative corrections to nuclear ground states
were extracted within $\chi$EFT.

Being interested in a larger set of observables, we present two alternative methods to the canonical perturbative 
higher-order calculation to avoid
accurate approximations of an infinite number of states. We formulate these methods for
any observable $\hat{O}$ and interaction $\hat{V}^{(0)}+\hat{V}^{(1)}$ for which approximate (or exact) solutions
can be obtained for either 
the wave function \emph{and} the energy or \emph{solely} the energy $E$ from 
\begin{equation}\label{eq.eveq}
\hat{H}\qty[\alpha,\beta] \ket{\psi} = 
    \qty(\hat{T}+\hat{V}^{(0)}+\alpha\cdot\hat{V}^{(1)}+\beta\cdot\hat{O})\ket{\psi}
    = E \ket{\psi}
\end{equation}
as functions of the real multipliers $\alpha$ and $\beta$ for the few-body system of interest.

In the following \secref{sec.meth}, we explain how to relate $\ket{\psi_n(\alpha,\beta)}$ and
$E_n(\alpha,\beta)$ to
\begin{equation}\label{eq.avg}
    \ev*{\hat{O}}_n
    = O_n^{(0)} + \delta O_n^{(1)} + \delta O_n^{(2)} + \dots\,,
\end{equation}
i.e., the perturbative corrections $\delta O_n^{(i)}$ proportional to the $i^{\text{th}}$ powers
of a small parameter $\alpha$. Section~\ref{sec.rad} reports an application of the methods to the point-proton
(root-)mean-squared charge radius, 
$\left<{r}_p^2\right>\stackrel{!}{=}\langle \psi|\hat{r}_p^2|\psi \rangle$, of the deuteron, triton, helion
and 4-helium ground states as
predicted with the non-perturbative leading order (LO) \nopi~interaction and its perturbative next-to-leading order (NLO) range correction \nopi~(see 
reviews~\cite{Van_Kolck_1999, Hammer:2019poc}). For the helion and 4-helium nuclei,
we analyze in \secref{sec.coul} \nopi~variants that treat the static 
Coulomb interaction as non-perturbative, perturbative, and negligible. We conclude in \secref{sec.pers}~with an
outlook on broader classes of problems to which the methodology can be applied.

\section{Obtaining corrections perturbatively}\label{sec.meth}
We present practical methods to obtain the perturbative expansion, \eqref{eq.avg}, without the need for an explicit infinite sum
as in \eqref{eq.psi1}. The summation is replaced by the requirement of a numerically complete basis, in which the eigenstate or 
eigenvalue in \eqref{eq.eveq}, relevant to the expectation value $\ev*{\hat{O}}$ of interest, can be expanded over a sufficient
range of multipliers $\alpha$ and $\beta$. The practicality of the methods comes with the availability of several sophisticated techniques developed for the quantum-mechanical few-body problem, e.g., lattice methods~\cite{Lee:2008fa,lattice2011}, 
Gaussian variational methods~\cite{Suzuki:1998bn,Silvestre-Brac:2007lvv}, 
Monte Carlo techniques~\cite{MCrev2015,MCrev2019}, 
Hyperspherical bases~\cite{nielsen2001three,Rittenhouse_2010}, Faddeev(-Yakubovsky) and AGS
decompositions~\cite{faddeev1960scattering,faddeev2016scattering,yakubovskii1967integral,Alt:1967fx}, 
resonating-group methods~\cite{wheeler1937molecular,Tang:1978zz,wildermuth2013unified}, 
and no core shell model \cite{ncsm2007,ncsm2013}. 
Any of these methods may prove useful in the determination of
the functional dependence of an expectation- or energy eigenvalue
in step $(I)$ of our methods. These methods entail nothing but a solution to Eq.~\eqref{eq.eveq} for the state of which an expectation value is of interest as a function of the single parameter $\alpha$ and an ensuing expansion of this dependence in a
power series. We refer to this method as the {\it one-multiplier method}~\eqref{eq.omm}:
\begin{equation}\label{eq.omm}
    \ev*{\hat{O}}_n
    \stackrel{(I)}{\equiv}
    O_n(\alpha;\beta=0)
    \stackrel{(II)}{=}
    O^{(0)}_n + 
    \sum_{i=1}^N
    \alpha^i ~\delta O^{(i)}_n~.
    \tag{OMM}
\end{equation}
While the coefficients of the power series furnish the perturbative 
expansion of the expectation value directly, a potential disadvantage of the approach is the demand of
a sufficiently accurate solution for an eigenstate over a range of $\alpha$'s, which must be chosen such that
the convergence radius of the expansion in $\alpha$ times the 
interaction strength of $\hat{V}^{(1)}$ is nonzero.

For cases where this latter issue arises or if no solution to the eigenvectors of \eqref{eq.eveq} is
available,
the {\it two-multiplier method}~\eqref{eq.tmm}, in turn, takes an intermediate step by
extracting solely eigenenergies of Eq.~\eqref{eq.eveq}. However, they must be obtained
on a two-dimensional $(\alpha,\beta)$ grid before adapting a
two-dimensional power series to the resultant energy contour:
\begin{equation}\label{eq.tmm}
    E_n
    \stackrel{(I)}{\equiv}
    E_n(\alpha,\beta)
    \stackrel{(II)}{=} 
    E_n^{(0,0)} +
    \sum_{\substack{i,j=0 \\ i \neq j =0}}^N
    \alpha^i\beta^j
    ~\delta E_n^{(i,j)}\;.
    \tag{TMM}
\end{equation}
As eigenenergies of the Schr\"{o}dinger equation~\eqref{eq.eveq}, the 
{\it Feynman-Hellmann theorem}~\cite{Feynman:1939zza,Hellmann:1937} applied with respect to the parameter $\beta$ 
\begin{equation}
\pdv{E_n(\alpha,\beta)}{\beta}=\ev{\pdv{\hat{H}\qty[\alpha,\beta]}{\beta}}_n = \ev*{\hat{O}}_n
\end{equation}
relates the expansion coefficients linear in $\beta$, $\delta E_n^{(i,1)}$, to the $i^{th}$ perturbation order of the 
observable:
\begin{equation}\label{eq.feynman-hellmann}
    O^{(0)}_n + 
    \sum_{i=1}^N
    \alpha^i ~\delta O^{(i)}_n
    =
    \delta E_n^{(0,1)} +
    \sum_{i =1}^N
    \alpha^i
    ~\delta E_n^{(i,1)}\,.
\end{equation}
The need to obtain eigenvalues $E_n(\alpha,\beta)$ on a two-dimensional grid significantly increases the computational cost
for a sufficiently robust two-parameter fit and renders the~\eqref{eq.tmm}~favorable only if the~\eqref{eq.omm}~turns out inadequate.

\section{Perturbative corrections to nuclear radii}\label{sec.rad}
As a first application of~\eqref{eq.omm}~and~\eqref{eq.tmm}~we investigate the point-proton root-mean-square charge radii (rms) of 
several small ($A\leq 4$) nuclei, namely, $^2$H, $^3$H, $^3$He, and $^4$He with a nuclear interaction theory that mandates the
perturbative treatment of some of its parts. More specifically, we use the potential formulation of the renormalizable~\nopi~up to 
NLO with local regulators~\cite{schafer2023few,bagnarol2023five,rojik2025charge,contessi2025renorm}. The LO interaction terms 
of this theory are iterated in the Schr\"{o}dinger equation, while NLO corrections must
be treated as a first-order
perturbation. In general, the higher-order perturbative corrections are expected to improve the theoretical accuracy 
of the predictions as more physics at higher momenta is systematically accounted for by the EFT. To employ the two numerical methods above for this specific interaction allows for 
benchmarking with earlier, alternative \nopi~perturbative extractions of~\rms~in $A\leq3$ systems: 
a semi-analytical approach for the deuteron~\cite{Chen1999} and two three-body 
nuclei investigated with the {\it dibaryon} formalism~\cite{Vanasse:2015fph,Vanasse:2017kgh}. 
No previous~\nopi~study of the point-proton root-mean-square 
charge radius beyond its non-perturbative LO is currently available for $A>3$, and our results for
$\Hef$ are thus first-time predictions within~\nopi.\\

Furthermore, we utilize the results to obtain insight into the interplay of a short-range (\nopi) and
long-range (static Coulomb) interactions. 
To this end, we 
adopt different potential incarnations of~\nopi, each of
which treating the static Coulomb interaction differently: 
as a LO and thereby a non-perturbative effect (cf. \eqref{pot_nonpC_lo} 
and \eqref{pot_nonpC_nlo} below, and for more details, see Ref.~\cite{rojik2025charge}); as a NLO perturbation (cf. 
\eqref{pot_pC_lo} and \eqref{pot_pC_nlo} below, and details in Ref.~\cite{contessi2025renorm}~); or entirely
absent~\cite{schafer2023few,bagnarol2023five}. For the first variant, namely, the \nopi~potential up to NLO with 
non-perturbative Coulomb, the interaction potential reads:

\begin{align}\label{pot_nonpC_lo}
    V^{(0)}_\text{C}=&
    \sum_{i<j}\left(C_0^{(0)}(\Lambda)\hat{\mathcal{P}}_{ij}^{(0,1;nn/np)}+C_1^{(0)}(\Lambda)\hat{\mathcal{P}}_{ij}^{(1,0;np)}\right)g_\Lambda(r_{ij})
    +\sum_{i<j}\left(C_2^{(0)}(\Lambda)\hat{\mathcal{P}}_{ij}^{(0,1;pp)}g_\Lambda(r_{ij}) + \frac{e^2}{r_{ij}}\hat{\mathcal{P}}_{ij}^{(pp)}\right)
    \nonumber\\
    &+\sum_{i<j<k}D_0^{(0)}(\Lambda)\sum_{\text{cyc}}\hat{\mathcal{P}}_{ijk}^{(1/2,1/2;nnp/ppn)}g_\Lambda(r_{ij})g_\Lambda(r_{ik})\,\qq{,}
    \\[10pt] \label{pot_nonpC_nlo}
    V^{(1)}_\text{C} =&
     \sum_{i<j}\left(C_0^{(1)}(\Lambda)~\hat{\mathcal{P}}_{ij}^{(0,1;nn/np)}+C_2^{(1)}(\Lambda)~\hat{\mathcal{P}}_{ij}^{(1,0;np)}
    +C_4^{(1)}(\Lambda)~\hat{\mathcal{P}}_{ij}^{(0,1;pp)}\right)g_\Lambda(r_{ij})
    \nonumber\\
    &+\sum_{i<j}\left(
     C_1^{(1)}(\Lambda)~\hat{\mathcal{P}}_{ij}^{(0,1;nn/np)}+C_3^{(1)}(\Lambda)~\hat{\mathcal{P}}_{ij}^{(1,0;np)}
    + C_5^{(1)}(\Lambda)~\hat{\mathcal{P}}_{ij}^{(0,1;pp)}\right)
    \left(g_\Lambda(r_{ij})\overset{\rightarrow}{\nabla}^2
        +\overset{\leftarrow}{\nabla}^2g_\Lambda(r_{ij})\right)
    \nonumber\\
    &+ \sum_{i<j<k} D_0^{(1)}(\Lambda)\sum_{\text{cyc}}\hat{\mathcal{P}}_{ijk}^{(1/2,1/2;nnp)}
    g_\Lambda(r_{ij})g_\Lambda(r_{ik})
    + \sum_{i<j<k} D_1^{(1)}(\Lambda)\sum_{\text{cyc}}\hat{\mathcal{P}}_{ijk}^{(1/2,1/2;ppn)} g_\Lambda(r_{ij})g_\Lambda(r_{ik})
    \nonumber\\
    &+ \sum_{i<j<k<l} E_{0}^{(1)}(\Lambda)~\hat{\mathcal{P}}^{(0,0)}_{ijkl}~g_\Lambda(r_{ijkl})\,\qq{.}
\end{align}

The $\hat{\mathcal{P}}^{({\rm S,T;ch})}_{\ldots}$ are generic projection operators into channels labeled by total spin ($S$), total isospin ($T$), and $T_z\longleftrightarrow$(ch), 
$g_\Lambda(r)=\text{exp}\left(-\frac{1}{4}\Lambda^2 r^2\right)$ are Gaussian regulators with momentum 
cutoffs $\Lambda$, $r_{ij} = \left| \pmb{r}_i - \pmb{r}_j \right|$ is a relative distance between particles $i$ and $j$, and $r^2_{ijkl} \equiv \displaystyle{\sum_{a<b\, \in\, \{i,j,k,l\}}} r_{ab}^2$ is the squared four-body hyperradius. For further details on the operator structure, the data used to calibrate the low-energy constants (LECs), and the numerical values of the latter, we refer the reader to Ref.~\cite{rojik2025charge}. The potential form of the second EFT variant, i.e.,
\nopi~up to NLO with perturbative Coulomb interaction, reads:

\begin{align}\label{pot_pC_lo}
    V^{(0)}_{\text C\!\!\!/}=&\sum_{i<j}\left(C_0^{(0)}(\Lambda)\hat{\mathcal{P}}_{ij}^{(0,1)}+C_1^{(0)}(\Lambda)\hat{\mathcal{P}}_{ij}^{(1,0)}\right)g_\Lambda(r_{ij})
    +\sum_{i<j<k}D_0^{(0)}(\Lambda)\sum_{\text{cyc}}\hat{\mathcal{P}}_{ijk}^{(1/2,1/2)}g_\Lambda(r_{ij})g_\Lambda(r_{ik})\,\qq{,}
    \\[10pt] \label{pot_pC_nlo}
    V^{(1)}_\text{pC} =& \sum_{i<j}\!\left(
        C_0^{(1)}(\Lambda)~\hat{\mathcal{P}}_{ij}^{(0,1)}
       +C_2^{(1)}(\Lambda)~\hat{\mathcal{P}}_{ij}^{(1,0)}
       \right)g_\Lambda(r_{ij})
    + \sum_{i<j}\!\left(
     C_1^{(1)}(\Lambda)~\hat{\mathcal{P}}_{ij}^{(0,1)}
    +C_3^{(1)}(\Lambda)~\hat{\mathcal{P}}_{ij}^{(1,0)}
    \right)\!\!
     \left(g_\Lambda(r_{ij})\overset{\rightarrow}{\nabla}^2
        +\overset{\leftarrow}{\nabla}^2g_\Lambda(r_{ij})\right)
    \nonumber\\
    &+\sum_{i<j}\left(C_{4;\text{pC}}^{(1)}(\Lambda)\hat{\mathcal{P}}_{ij}^{(0,1;pp)}g_\Lambda(r_{ij}) + \frac{e^2}{r_{ij}}\hat{\mathcal{P}}_{ij}^{(pp)}\right)
    + \sum_{i<j<k} D_0^{(1)}(\Lambda)\sum_{\text{cyc}}\hat{\mathcal{P}}_{ijk}^{(1/2,1/2)}
    g_\Lambda(r_{ij})g_\Lambda(r_{ik})
    \nonumber\\
    &+ \sum_{i<j<k<l} E_{0;\text{pC}}^{(1)}(\Lambda)
    ~\hat{\mathcal{P}}^{(0,0)}_{ijkl}~g_\Lambda(r_{ijkl})\qq{.}
\end{align}

Note the isospin symmetry of the LO potential, $V^{(0)}_\text{\text C\!\!\!/}$. Thence, it does not differentiate between 
projections into the third isospin axis ($nn/np/pp$). The Coulomb interaction is accounted for perturbatively as a part of the
NLO range correction, $V^{(1)}_\text{pC}$. 

The third EFT version does not consider the static Coulomb interaction even at NLO ($V^{(1)}_\text{\text C\!\!\!/}$). 
Therefore, its form derives from~\eqref{pot_pC_nlo}~by setting the $C_{4;\text{pC}}^{(1)}$ and 
Coulomb terms in $V^{(1)}_\text{pC}$ to zero. The corresponding four-body-force LEC, $E_{0;\text{\text C\!\!\!/}}^{(1)}$, is
always readjusted such that the ground-state binding energy of $^4$He is reproduced. 

In the course of renormalizing these three EFTs, i.e. in practice, fitting their LECs,
we use a single set of data. 
For $C_{4;\text{pC}}^{(1)}$ of Eq.~\eqref{pot_nonpC_nlo}, in particular, we fit to the
ground-state binding-energy difference $B(^3\text{H})-B(^3\text{He})$ (see Ref.~\cite{contessi2025renorm}).

Having thus split the interacting theory into a non-perturbative and a perturbative part - ($V^{(0)}_\text{C} + V^{(1)}_\text{C}$), 
($V^{(0)}_\text{\text C\!\!\!/} + V^{(1)}_\text{pC}$), and ($V^{(0)}_\text{\text C\!\!\!/} + V^{(1)}_\text{\text C\!\!\!/}$), the resulting Eq.~\eqref{eq.eveq}~must be solved for the various nuclear ground states on a one-dimensional $\alpha$ 
grid~\eqref{eq.omm}~and for the ground-state energies on a two-dimensional ($\alpha,\beta$) grid~\eqref{eq.tmm}. To do so, we
use the Stochastic Variational Method with correlated Gaussian basis \cite{Suzuki:1998bn} for sets $\left\lbrace
    \ev*{{r}^2_p}_0(\alpha; \beta=0),~\alpha\in I    
    \right\rbrace$ and $\left\lbrace
    E_0\qty(\alpha,\beta),~\qty(\alpha\cdot\beta)\in I\otimes I    
    \right\rbrace$ on the logarithmic grid
\begin{equation}\label{eq.grd}
    I
    =
    \left\lbrace
    e^{\ln(x_1)+\qty[\ln(x_{M})-\ln(x_{1})]\cdot n/M}
    ~,~n\in\qty[0,\ldots,M]    
    \right\rbrace \qq{,}
\end{equation}
with $M=25$, $x_1=10^{-10}$, and $x_{M}=3\cdot10^{-3}$. Here, each element (26 in case of the~\eqref{eq.omm}~and $26^2$ 
for~\eqref{eq.tmm}) is the result of an independent variational solution of Eq.~\eqref{eq.eveq} for the approximation of the respective nuclear ground states. For each $\Lambda$, we diagonalize within one larger correlated-Gaussian basis, stochastically selected with the aid of auxiliary harmonic oscillator traps \cite{schafer2021}. The subsequent least-square~\eqref{eq.omm}~and~\eqref{eq.tmm}~fit
in step $(II)$ is employed with an increasing polynomial order up to the point where the numerical stabilization of extracted 
NLO perturbative corrections, $\delta O^{(1)}_0 = \ev*{{r}^2_p}_0^{(1)}$, Eq.~\eqref{eq.avg}, is achieved.

\subsection{Benchmarking in $A\leq4$ nuclei}

For \nopi~up to NLO with non-perturbative Coulomb interaction, Eq.~\eqref{pot_nonpC_nlo}, we calculate point-proton charge 
root-mean-square radii of light nuclei at different cutoff values $\Lambda\in\lbrace 1,2,\dots,10\rbrace\,\text{fm}^{-1}$. Both 
\eqref{eq.omm}~and~\eqref{eq.tmm} methods are applied to extract the non-perturbative LO values, $\ev*{{r}^2_p}_0^{(0)}$--in the notation of Eq.~\eqref{eq.avg}, the subscript $(0)$ marks the ground state and the superscript the perturbation order-- and perturbative NLO corrections, $\delta \ev*{{r}^2_p}_0^{(1)}$, to them.
The EFT prediction at lowest order for the square root of an observable is
\begin{equation}
\sqrt{\ev*{{r}^2_p}_0}\Bigg|_\text{LO} \equiv  \sqrt{\ev*{{r}^2_p}_0^{(0)}}
\qq{,}    
\end{equation}
while the square root must be expanded at NLO as
\begin{equation}
\sqrt{\ev*{{r}^2_p}_0}\Bigg|_\text{NLO} =  \sqrt{\ev*{{r}^2_p}_0}\Bigg|_\text{LO} 
\left(1 + \frac{\delta \ev*{{r}^2_p}_0^{(1)}}{2 \ev*{{r}^2_p}_0^{(0)}}\right)\qq{,}
\end{equation}
to have a polynomial order consistent with the EFT.

\begin{figure*}
\begin{tabular}{cc}
\includegraphics[width=0.48\columnwidth]{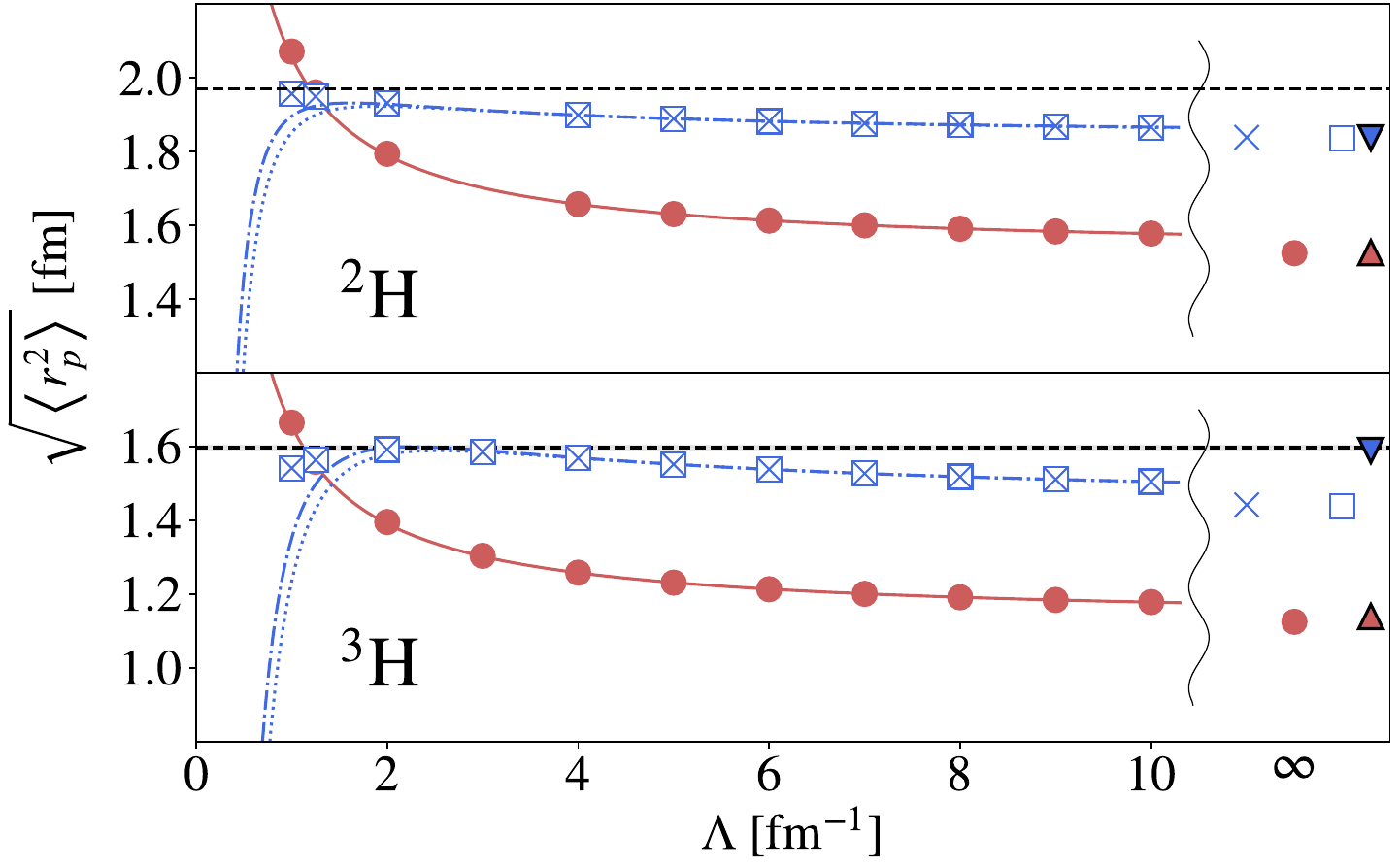}& \includegraphics[width=0.48\columnwidth]{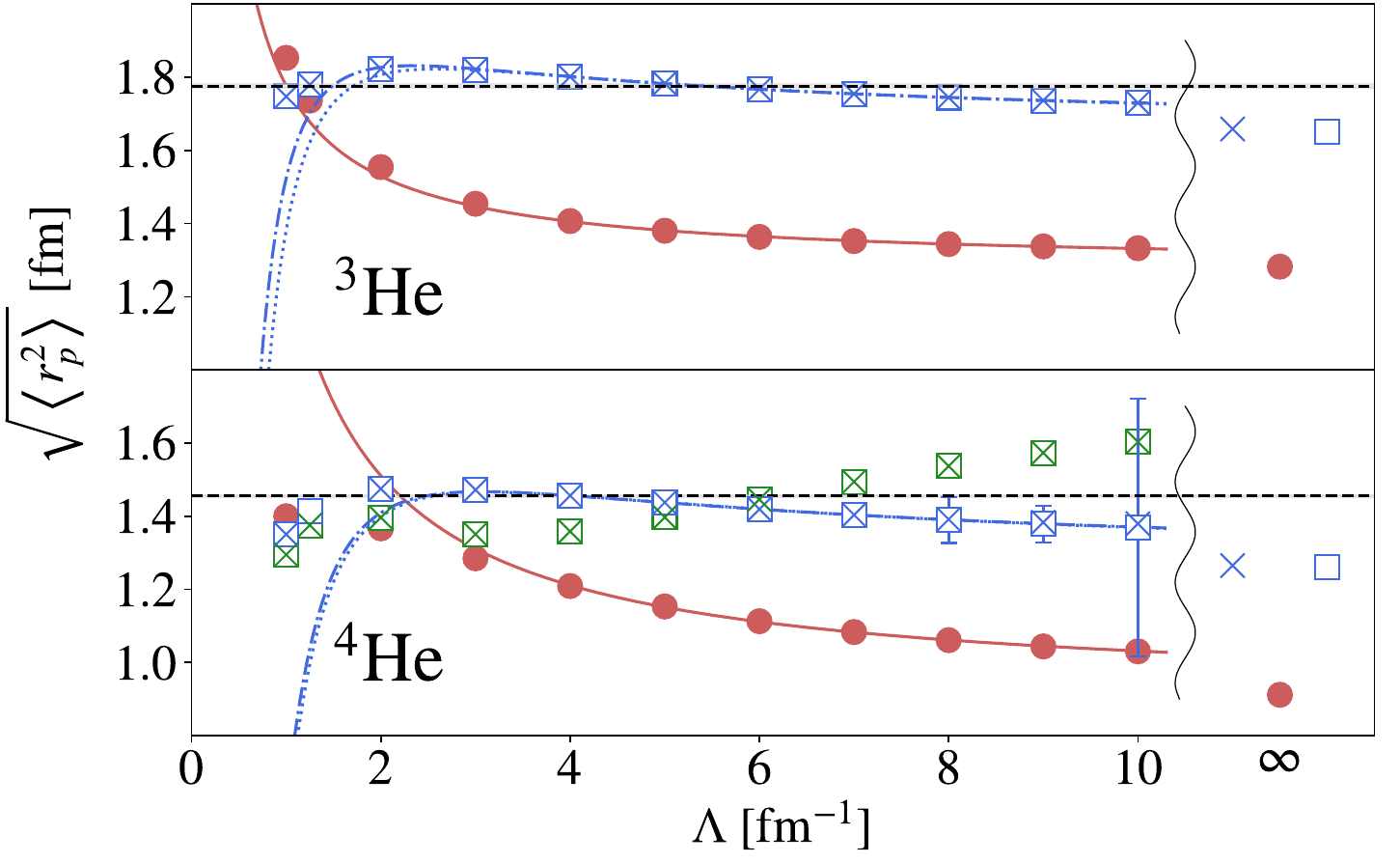}\\
\end{tabular}
\caption{Cutoff regulator ($\Lambda$) dependence of \rms~predictions at LO
\raisebox{-2pt}{
\protect\tikz{
\protect\draw [thick, red] 
    (0,0) 
    -- (0.5,0)
    node [red, xshift=.0cm, yshift=-.0cm,midway] {\protect\tikz{\protect\pgfuseplotmark{*}}}
;
}}
and NLO [\eqref{eq.omm}
\raisebox{-2pt}{
\protect\tikz{
\protect\draw [thick, blue, dotted] 
    (0,0) 
    -- (0.55,0)
    node [blue, xshift=.0cm, yshift=-.0cm,midway] {\protect\tikz{\protect\pgfuseplotmark{square}}}
;
}}
, \eqref{eq.tmm}
\raisebox{-4pt}{
\protect\tikz{
\protect\draw [thick, blue, dash dot] 
    (0,0) 
    -- (0.65,0)
    node [blue, xshift=.0cm, yshift=-.0cm,midway,scale=1.] {$\times$}
;
}}
] of \nopi~for $A\leq4$ nuclei. 
$\Lambda\to\infty$ limits are split by
\raisebox{-1pt}{\protect\tikz{\protect\path [draw=black,snake it,scale=0.05, segment length=4] (0,1.5) -- (0,-4.5);}}
and compared for \Hd~with~
\cite{Chen1999}~(LO
\raisebox{1pt}{\protect\tikz{\protect\node[thick,draw,scale=0.3,regular polygon, regular polygon sides=3,fill=red](){};}}
,
NLO
\raisebox{1pt}{\protect\tikz{\protect\node[thick,draw,scale=0.3,regular polygon, regular polygon sides=3,fill=blue,rotate=180](){};}}
)
and for \Ht~with~\cite{Vanasse:2015fph}~(LO
\raisebox{1pt}{\protect\tikz{\protect\node[thick,draw,scale=0.3,regular polygon, regular polygon sides=3,fill=red](){};}}
, 
NLO
\raisebox{1pt}{\protect\tikz{\protect\node[thick,draw,scale=0.3,regular polygon, regular polygon sides=3,fill=blue,rotate=180](){};}}
). For \Hef, we also show results without the four-body counter-term at NLO
[\eqref{eq.omm} \raisebox{1pt}{\protect\tikz{\protect\node[draw,green,scale=0.5,regular polygon, regular polygon sides=4,fill=white](){};}}\,,
\eqref{eq.tmm}\raisebox{-3pt}{\protect\tikz{\protect\node[green](){$\times$};}}]. The Coulomb interaction is considered non-perturbatively. 
All results are shown next to experimental data (\raisebox{2pt}{\protect\tikz{\protect\draw[dashed,thick](0,0)--(0.5,0);}}), which are 
evaluated as $\sqrt{\left<r^2_p \right>} = \sqrt{\left<r^2_p \right>_{ch}-\left<R^2_p \right>-\frac{N}{Z}\left<R^2_n \right>_{ch}-\frac{3}{4 M_p^2}}$. 
We take experimental charge mean-squared radii $\left<r^2_p \right>_{ch}$, from Ref.~\cite{ANGELI201369}, proton/neutron mean-squared 
charge radii $\left<R^2_p \right>$/$\left<R^2_n \right>$ from Ref.~\cite{pdg2024}, and a value of Darwin-Foldy term $3/(4 M_p^2)=0.033~\text{fm}^2$.}
\label{fig.respC}
\end{figure*}

In \figref{fig.respC}, we display calculated $^2$H, $^3$H, $^3$He, and $^4$He ground state radii at LO and NLO
as functions of an increasing momentum cutoff $\Lambda$. At NLO, 
we find that NLO corrections calculated with \eqref{eq.omm}~and~\eqref{eq.tmm} differ by $\lesssim 0.1\%$
except for $^4$He values at higher cutoff values extracted using \eqref{eq.tmm}. 
For the latter, we obtain relatively large uncertainties from the fits of 
the polynomials to the calculated energy grids, \eqref{eq.grd}. We attribute this effect to a limited flexibility of the
correlated Gaussian basis implying an increasing difficulty to attain numerical convergence at larger cutoff values. In contrast, \eqref{eq.omm} maintains its 
numerical stability, and fitting errors are smaller than the resolution of the figure for all results. 

Since we benchmark our results against \nopi, we detail the $\Lambda\to\infty$ extrapolation of our finite-cutoff results. We 
assume
\begin{equation}\label{eq.rescutdep}
    \ev{{r}^2_p}\qty(\text{N}^{m}\text{LO})
    =
    \sum_{n=0}^{m+1}
    \frac{b^{(m)}_{n}}{\Lambda^{n}}\,\qq{,}
\end{equation}
for the residual dependence of all radii on the Gaussian regulator parameter $\Lambda$ at order $m$ of the EFT. Thus, we fit two parameters ($b^{(0)}_{0}$ and $b^{(0)}_{1}$) at LO and three ($b^{(1)}_{0}$, $b^{(1)}_{1}$, and $b^{(1)}_{2}$) at NLO.

In the upper-left panel of \figref{fig.respC}, we compare our results for~\rms~as predicted by the LO and NLO of \nopi~for the deuteron with the corresponding results calculated in Ref.~\cite{Chen1999}. The observable, interaction structure, and the data 
used to calibrate its strengths are identical. However, the earlier work obtains the leading perturbative correction within the
semi-analytical framework of perturbative EFT. This prediction starts from
the same~\nopi~Lagrangian density of which the
potential operators used in Eq.~\eqref{eq.eveq} are derived. A noteworthy difference, however, is their use of a power-divergence 
subtraction scheme~\cite{Chen1999} in contrast to the cutoff regularization employed in this work. It is therefore a non-trivial
observation and consistency check that, in the zero-range limit $\Lambda\to\infty$, our results for the non-perturbative and 
perturbative orders yield the same numerical result.

Given this agreement for the deuteron, 
our NLO \rms\, results for $\Ht$ (lower-left panel 
of \figref{fig.respC}) extrapolated to the contact limit differ from those of Ref.~\cite{Vanasse:2015fph}: 
$0.9\%$ at LO and $9\%$ at NLO. This hints towards an increase of the NLO over the LO contribution in the three-body case
compared with the two-body system.
A plausible explanation for this difference\footnote{Note that a naive (10)30\%~estimate of the theoretical (N)LO uncertainty
in the \nopi~prediction does cover the observed difference.}, and especially its increase, when comparing LO to NLO between 
the two works, is the employed renormalization condition. The agreement between the LO and NLO deuteron radii arises when both the 
diagrammatic and our numerical approaches renormalize the interaction strength to an amplitude that is expanded in the effective
range of the spin-triplet neutron-proton system (the so-called $\rho$-parametrization). In contrast, the $^3$H calculation in 
Ref.~\cite{Vanasse:2015fph}~chose to renormalize the two-nucleon propagators at NLO by fixing the residues of the bound-state and
virtual poles in the two (iso)spin channels. This approach (so-called the $Z$-parametrization) has been 
developed in Ref.~\cite{Phillips_2000,Griesshammer:2004pe}~to improve the convergence of \nopi, especially for observables which
are sensitive to external probes that induce (virtual) transitions from a bound to a scattering state. The operator $\hat{r}^2_p$ 
falls into this category, and our results are therefore expected to converge more slowly to the experimental $\Ht$ datum. We intend
to revisit this explanation in future work once the $Z$-parametrized \nopi~potential is available. Furthermore, we expect a less
pronounced effect when switching from $\rho$- to $Z$-parametrization in $\Hef$~as its ground state is far below the deuteron-deuteron 
threshold.

Furthermore, the earlier \nopi~studies of the $\Het$ point-proton root-mean-square charge radius
\cite{Vanasse:2017kgh} represent a qualitatively 
different theory as they treat the proton-proton system without any electromagnetic contribution. Our results (upper-right panel of \figref{fig.respC}), in turn,
employ a non-perturbative iteration of the static Coulomb potential. An 
intuitive argument suggests, however, a perturbative character of the Coulomb force for sufficiently large momenta. Whether this 
justifies its demotion or complete nullification in the study of bound systems relates to the question of the characteristic momenta
at which charges approach each other within those compounds. 
Naively, the deeper the binding, the larger the typical relative binding 
momentum, and the more perturbative the Coulomb interaction becomes. 
We revisit this issue of `perturbativeness' of the Coulomb 
interaction in the following subsection.

To date, no other assessment exists of the effect of a perturbative NLO range correction on \rms~of $\Hef$.
At LO, our
results agree (lower-right panel in \figref{fig.respC}) with an earlier \nopi~Monte-Carlo calculation \cite{Contessi:2017rww} which entails a remarkable consistency in regulator-parameter convergencies, $\Lambda\to\infty$, of equal accuracy as those
observed in the LO three-body calculations. 
In contrast to the three-nucleon system, $\Hef$ is characterized by a new four-body scale which is probed with NLO 
operators~\cite{Bazak:2018qnu}. Leaving this scale unconstrained can enhance the regulator dependence to a degree that may render the 
theory impractical (green markers - \eqref{eq.omm}\raisebox{-3pt}{\protect\tikz{\protect\node[green](){$\times$};}} and 
\eqref{eq.tmm}~\raisebox{1pt}{\protect\tikz{\protect\node[draw,green,scale=0.5,regular polygon, regular polygon 
sides=4,fill=white](){};}}). This demonstrates, for the first time, the effect of the NLO deviation from naive power counting in a 
four-body observable other than the binding energy.~(cf. Ref.~\cite{Bazak:2018qnu}).

\subsection{Accounting for the Coulomb Interaction}\label{sec.coul}
For another non-trivial application of the introduced perturbative tools, we chose to investigate differences in \rms~predictions at LO 
and NLO of the earlier introduced static-Coulomb versions of \nopi. This provides a quantitative guide to the circumstances, namely, the observables, for which one formulation converges order-by-order more rapidly than the other. In technical terms, we compare the effects of the absent \mbox{($V^{(0)}_\text{\text C\!\!\!/} + V^{(1)}_\text{\text C\!\!\!/}$),} perturbative 
\mbox{($V^{(0)}_\text{\text C\!\!\!/} + V^{(1)}_\text{pC}$),} and non-perturbative \mbox{($V^{(0)}_\text{C} + V^{(1)}_\text{C}$)}
static-Coulomb interaction on \rms~of \Het~and \Hef. The corresponding LO and NLO predictions in their scaling limits ($\Lambda\to\infty$)
are compiled\footnote{The adequacy of the \ref{eq.omm} was assessed in $^4$He for the non-perturbative Coulomb EFT and we subsequently abstained from obtaining the computationally more costly alternative \ref{eq.tmm} value for the other cases.} in~\tabref{tab.radii}.

\begin{table}[t!]
\centering
    \addtolength{\tabcolsep}{-1pt}    
\begin{tabular}{ccccccc}
\hline\hline
\multirow{2}{*}{${}^{A}X$}& {\scriptsize Coulomb} & \multirow{2}{*}{$\Lambda^*$} & 
\multirow{2}{*}{$\pi\!\!\!/_{\text{\tiny LO}}$} & $\pi\!\!\!/_{\text{\tiny NLO}}$
& $\pi\!\!\!/_{\text{\tiny NLO}}$ & Exp.\\
&  {\scriptsize versions}       &             &                                &  \eqref{eq.omm}     
&  \eqref{eq.tmm}    &   $\sqrt{\left<r^2_p \right>}$    \\
\hline\hline 
$^2$H& -  &         -     & 1.52  & 1.84 & 1.84 & 1.990 \\[10pt]
$^3$H& -  & $B\qty(\Ht)$  & 1.13  & 1.45 & 1.44 & 1.608 \\[10pt]
$^3$He& \text C\!\!\!/  & $B\qty(\Het)$ & 1.33  & 1.69 & 1.68 & \multirow{ 4}{*}{1.784}\\[2pt]
& \text C\!\!\!/  & $B\qty(\Ht)$  & 1.25  & 1.60 & 1.60 & \\[2pt]
& pC  & $B\qty(\Ht)$  & 1.25  & 1.63 & 1.63 &  \\[2pt]
& C & $B\qty(\Ht)$  & 1.28  & 1.66 & 1.65 &   \\[10pt]
$^4$He& \text C\!\!\!/  & $B\qty(\Het)$ & 0.961 & 1.24 & -   &\multirow{ 4}{*}{1.477}\\[2pt]
       & \text C\!\!\!/  & $B\qty(\Ht)$  & 0.905 & 1.30 & -    &   \\[2pt]
       & pC  & $B\qty(\Ht)$  & 0.905 & 1.24 & -    &   \\[2pt]
       & C & $B\qty(\Ht)$  & 0.912 & 1.27 & 1.26 &  \\
       \hline\hline
\end{tabular}
\caption{Predictions for various nuclear ($A\leq 4$) point-proton root-mean-square charge radii \rms~(in fm) of 
\nopi~versions differing in their Coulomb treatment [absent (\text C\!\!\!/)/perturbative
(pC)/non-perturbative (C)] and three-body scale input $\Lambda^*$. All numbers are 
considered in the scaling limit, $\Lambda\to\infty$, and the associated extrapolation
uncertainty is $<10^{-3}$~fm. For references to experimental data, see the caption of 
\figref{fig.respC}.}
\label{tab.radii}
\end{table}

First, we observe that isospin-symmetric {\text C\!\!\!/~} predicts a smaller \Het~radius if the binding energy of the triton sets the 
three-body scale, while it gets larger using the helion binding energy. The results are thereby consistent
with the naively expected negative-slope 
correlation between binding energy and radius.
For \Hef, the {\text C\!\!\!/~} LO root-mean-square radius prediction is found larger, too, if
the shallower $B\qty(\Het)$ scale is employed instead of the deeper $B\qty(\Ht)$ as renormalization condition.
However, at NLO, the radius is predicted to be larger with the latter constraint  --  an effect which has to be related to the new four-body NLO scale
and the chosen calibration of the same $B\qty(\Hef)$ datum for both three-body scenarios.

The perturbative consideration of the Coulomb exchange has, by construction, no effect at LO, and is of repulsive character in \Het.
There, it induces 
an increase from $1.6\,$fm to $1.63\,$fm. 
For \Hef, the Coulomb perturbation seemingly shrinks the system from $1.3\,$fm to $1.24\,$fm,
while the four-body counter-term enforces the same binding energy
as in the \text C\!\!\!/ EFT. 
Comparing the pC and fully-iterated C EFTs, the latter predicts larger \rms~of \Het~and \Hef~nuclei at LO and NLO relative to the former.

An important indicator of the convergence rate of the respective EFTs is the
difference between LO and NLO predictions. In this regard, the three EFT variants do not differ significantly; all implying a rate
consistent with the na\"ive $1/3$ \nopi~estimate. 
Quantitatively, the Coulomb treatment affects the NLO contribution to vary by $\lesssim0.03$~fm for 
\Het, and $\lesssim 0.07$~fm for \Hef, while the \nopi~NLO uncertainty estimates based on cutoff variation including the
$\Lambda \rightarrow \infty$ limit are, respectively, $\approx0.16$~fm and $0.2$~fm.

In our study, the \nopi's theoretical uncertainty is caused by a truncation of a double-expansion in two small parameters--$\left(p_\text{ty}/m_\pi\right)$ 
associated with a contact interaction whose perturbation series breaks down at large typical momenta, $p_\text{ty}$, but converges for lower momenta, and $\left(\alpha m_N/p_\text{ty}\right)$ characterizing a long-range Coulomb interaction where the power series diverges at small momenta. The observation of the \nopi~being dominated for a specific observable by one of those expansions is the key result. From our predictions, we conclude that the ground-state root-mean-square radius is insensitive 
to long-range Coulomb effects up to NLO in \nopi. The momenta contributing significantly to the ground state radii appear 
sufficiently large to suppress static Coulomb exchange relative to the short-distance interaction.

We are thus led to conclude that observables on $A$-body nuclear states with a typical momentum scale, $p_\text{ty}^{(A)}$,
-- which may be associated with energies well below some breakup threshold, e.g., in our case, we have
\begin{equation}
\frac{p_\text{ty}^{(3)}}{p_\text{ty}^{(4)}}:=\frac{\sqrt{\frac{2}{3}m_NB\qty(\Het)}}{\sqrt{\frac{3}{4}m_N B\qty(\Hef)}}\approx \frac{1}{3}
\end{equation}
-- of the order considered here can be accurately predicted without having to deal with the
long-range part of the interaction, at all. To see significant differences in the
predictions of the three introduced EFTs, it is now obvious to turn to observables on states
which involve much lower typical momenta, e.g., radii of halo nuclei, or transitions between the
ground and shallow nuclear states, such as the first excited state of \Hef.

\section{Perspectives}\label{sec.pers}
The methods described in this article extract from numerical solutions of the Schr{\"o}dinger equation corrections to observables 
expressed as polynomials of arbitrary order in a general perturbation operator. The efficacy of the methods 
was demonstrated by assessing the perturbative effect of the long- and short-range components of a nuclear 
interaction theory on the second spatial moment of the proton distribution within the nuclear bound states 
comprising less than five nucleons.

The renormalizable NLO \nopi{} prediction of the \Hef~point-proton root-mean-square charge radius, together with our quantification of the impact of excluding the Coulomb interaction, treating it in first-order perturbation theory, and iterating it to all orders on \rms~of \Het~and~\Hef~nuclei, are results original to our work.

The application of the methods to calibrate the coupling strengths of \nopi~beyond NLO 
renders them stepping stones to assess with these higher orders usefulness and convergence
rates of \nopi~for different observables.
Furthermore, we deem the methods practical to analyze the ``perturbativeness" of certain parts of
other nuclear interaction potentials. 
For instance, the perturbative character of an exchange pion or $\Delta$-excitation as part of 
high-precision interactions (e.g.,~\cite{Epelbaum_2009,Piarulli_2016,Somasundaram_2024}) can 
readily be studied in analogy to our above assessment of the Coulomb interaction. 
Finally, any prediction of scattering observables that extracts amplitudes from 
expectation values, e.g., via the finite-volume or trap-size dependence of energy spectra, 
can be combined with our methods. The associated perturbation-induced transformation of the character 
of $S$-matrix poles--a controversial issue in the development of contact 
theories (see~\cite{Sch_fer_2021,Yang_2024})--from, e.g., bound$\to$virtual/resonance, or vice-versa, is thus transparent:
the pole moves as a consequence of an infinite iteration, and it is from this infinite series one isolates the various orders.

\section*{Acknowledgments}
S.M. acknowledges financial support from the Department of Science and Technology (DST), Government of India (GOI), under the INSPIRE scheme (Fellowship number IF190758). He also thanks Soumyadeep Mondal for valuable discussions during the initial stage of the theoretical development. In addition, S.M. gratefully acknowledges SRM University, AP, India, for providing accommodation and access to office facilities, where the theoretical part of this work was primarily developed. The work of M.S. was supported by the Czech Science Foundation GACR grant 25-15746O. M.B. and N.B. acknowledge the Israel Science Foundation for its support under the grant ISF 2441/24. U.R. acknowledges financial support from the Science and Engineering Research Board (SERB), Government of India [MATRICS scheme grant number MTR/2022/000067 and CORE RESEARCH scheme grant number CRG/2022/000027]. He further acknowledges travel support provided under these grants for J.K. and M.S. during their academic visits to IIT Guwahati.

\bibliographystyle{unsrt}
\bibliography{rmsNLO.bib}

\end{document}